# Numerical Simulation of Magnetic Interactions in Polycrystalline YFeO$_3$

**E. Lima Jr.[1], T. B. Martins[1], H. R. Rechenberg[1], G. F. Goya[1],**

**C. Cavelius[2], R. Rapalaviciute[2], S. Hao[2], S. Mathur[2]**

[1] Instituto de Física, Universidade de São Paulo CP 66318, 05315-970, São Paulo, Brazil.

[2] Leibniz Institute of New Materials, CVD Division, D-66123 Saarbrücken, Germany.

**ABSTRACT**

The magnetic behavior of polycrystalline yttrium orthoferrite was studied from the experimental and theoretical points of view. Magnetization measurements up to 170 kOe were carried out on a single-phase YFeO$_3$ sample synthesized from heterobimetallic alkoxides. The complex interplay between weak-ferromagnetic and antiferromagnetic interactions, observed in the experimental $M(H)$ curves, was successfully simulated by locally minimizing the magnetic energy of two interacting Fe sublattices. The resulting values of exchange field ($H_E$ = 5590 kOe), anisotropy field ($H_A$ = 0.5 kOe) and Dzyaloshinsky-Moriya antisymmetric field ($H_D$ = 149 kOe) are in good agreement with previous reports on this system.

## I. INTRODUCTION

Orthoferrites of formula $REFeO_3$, in which RE is a rare earth element, have remarkable magnetic properties of primary significance for technological applications [1-3]. Although the crystal structure of these oxides is essentially insensitive to the RE element, the corresponding magnetic properties can differ markedly depending on the single-ion magnetic character of the rare earth involved. Indeed, when RE is a magnetic ion, the existence of two magnetic sublattices hinders the study of exchange coupling within the Fe sublattice, therefore nonmagnetic Y has been chosen for this purpose.

The $YFeO_3$ orthoferrite crystallizes in the $D_{2h}^{16}$ – $Pnma$ space group, with a distorted perovskite structure [4]. In this structure there is only one $Fe^{3+}$ site, octahedrally coordinated to the six nearest oxygen ions [5]. Below $T_N \sim 650$ K the system is antiferromagnetically ordered, with a slight canting of Fe spins giving rise to a weak-ferromagnetic (WFM) moment that dominates the magnetic behavior of the system. The origin of the canting is the antisymmetric exchange Dzyaloshinsky-Moriya (DM) interaction $-\vec{D} \cdot (\vec{M}_1 \times \vec{M}_2)$ where $\vec{D}$ is the DM vector and $\vec{M}_1$, $\vec{M}_2$ are the sublattice magnetization vectors [6]. In addition, this system also presents a uniaxial magnetocrystalline anisotropy in a direction perpendicular to the DM vector. Thus, an expression for magnetic energy including all relevant terms to describe the magnetic behavior of the $YFeO_3$ system within a two-sublattice approximation is given by [7-9]:

$$U = \lambda(\vec{M}_1 \cdot \vec{M}_2) - \vec{H} \cdot (\vec{M}_1 + \vec{M}_2) - \vec{D} \cdot (\vec{M}_1 \times \vec{M}_2) - \left(\frac{K_2}{M_0^2}\right)(M_{1x}^2 + M_{2x}^2) \quad (1)$$

where $M_1 = M_2 = M_0$ and the **x** direction is taken as the easy magnetization axis. The first term in Eq. 1 is the isotropic exchange interaction that forces the magnetic moments to align antiparallel; the second term represents the interaction of magnetic moments with the applied field; the third term is the DM interaction that aligns the vectors $\vec{D}$, $\vec{M}_1$ and $\vec{M}_2$ perpendicular to each other; and the last term represents the anisotropy energy.

The distinctive magnetic structure of YFeO$_3$ yields specific magnetic properties originating from the coexistence of small and large anisotropy fields associated to magnetocrystalline and DM interactions, respectively. In this way, this system exhibits unusual magnetization curves as a function of applied field and temperature. In particular, the hysteresis loop measured up to fields below the irreversibility field ($H_{irr}$ ~ 80 kOe, see next section) can exhibit low- and high-field regimes associated to the small and large anisotropy fields, respectively. These two field regimes have been attributed [10] to a mixture of hard and soft magnetic phases in the sample; indeed, it is known from the Y-Fe-O phase diagram that YFeO$_3$ samples prepared through solid-state reactions of other physical methods usually contain undesired binary (Fe$_3$O$_4$) and ternary (Y$_3$Fe$_5$O$_{12}$) secondary phases. However, in a recent paper Mathur *et al.* have reported [11] the synthesis of nanocrystalline YFeO$_3$ from Y-Fe mixed-metal alkoxide [YFe(OPr$^i$)$_6$(Pr$^i$OH)]. The method was successful in producing single-phase samples of highly stoichiometric and homogeneous composition, at remarkably low temperatures. In spite of the improved sample quality, the low- and high-field regimes were also apparent in magnetization loops measured for fields up to 70 kOe, which could not be attributed to a phase mixture. As an example, Fig. 1 shows minor hysteresis loops of a sample prepared as described in Ref. 11 and calcined at 1173 K. This two-regime behavior was attributed [11] to the combination/competition of magnetocrystalline and DM interactions, as an intrinsic characteristic of YFeO$_3$.

FIGURE 1

The present paper reports a detailed theoretical study of the origin of the observed peculiarities in the magnetization curves of a single-phase, polycrystalline sample of $YFeO_3$ prepared by the above-mentioned method. Magnetization loops were calculated using a local energy minimum approach and the results were very close to the experimental curves measured up to $H = 170$ kOe, yielding $H_E$, $H_A$ and $H_D$ values in good agreement with those reported in the literature for this system.

## II. EXPERIMENTAL

A pure and well-crystallized $YFeO_3$ powder sample with sub-micrometric grain size (~ 300-500 nm) was prepared by sol-gel processing of the Y-Fe mixed metal alkoxide [$YFe(OPr^i)_6(Pr^iOH)$] and calcined at 1173 K as previously described by Mathur et. al. Single-phase character and high crystallinity were evidenced from x-ray diffraction and Mössbauer spectroscopy [11]. Fig. 2 exhibits a magnetization loop of this sample measured in a vibrating-sample magnetometer at 4.2 K, where only the high-field regime is present. The sample shows a high coercive field $H_c \sim 20$ kOe and a high irreversibility field $H_{irr} \sim 80$ kOe. $H_{irr}$ is defined as the threshold field above which the magnetic energy of the system presents a single energy minimum, indicated by the disappearance of hysteresis in the magnetization loop. Magnetization at the largest measuring field $H = 170$ kOe is only 2.3 emu/g, far off the saturation value $M_s = 72$ emu/g expected for a 5 $\mu_B$ moment per $Fe^{3+}$ ion. The latter result is direct evidence for a very large exchange field, preventing saturation to be attained in available experimental fields; this feature will be further examined in the following section.

FIGURE 2

## III. SINGLE-CRYSTALLINE SIMULATIONS

Fig. 3 shows a schematic diagram of the vectors $\vec{H}$, $\vec{D}$, $\vec{M}_1$ and $\vec{M}_2$ involved in the calculation, following the convention introduced in Ref. 12. The directions of vectors $\vec{H}$, $\vec{M}_1$ and $\vec{M}_2$ are specified by polar angles ($\gamma$, $\delta$), ($\alpha$, $\theta$) and ($\beta$, $\varphi$) respectively; vector $\vec{D}$ lies along the **y** axis and the anisotropy axis is along the **x** direction. Consequently, the WFM component lies along the **z** axis in zero applied field. $\vec{M}_1$ and $\vec{M}_2$ have the same magnitude $M_0$, taken as 36 emu/g. In order to write Eq. 1 in terms of interaction fields, one defines the exchange field $H_E = \lambda M_0$, the uniaxial anisotropy field $H_A = 2K/M_0$ and the DM field $H_D = DM_0$, thus the magnetic energy density $u = U/M_0$ can be expressed as:

$$u = H_E[\sin\alpha \sin\beta \cos(\theta - \varphi) + \cos\alpha \cos\beta]$$
$$+ H\{\sin\gamma[\sin\alpha \cos(\theta - \delta) + \sin\beta \cos(\varphi - \delta)] + \cos\gamma[\cos\alpha + \cos\beta]\}$$
$$+ \frac{H_A}{2}(\sin^2\alpha + \sin^2\beta) + H_D(\sin\alpha \cos\beta \sin\theta - \cos\alpha \sin\beta \sin\varphi) \quad (2)$$

FIGURE 3

For a given field $\vec{H}$, the equilibrium configuration is obtained by minimizing Eq. 2 with respect to angles ($\alpha$, $\beta$, $\theta$ and $\varphi$) and the normalized magnetization component along the applied field is calculated through the following expression:

$$m = \frac{\vec{M} \cdot \vec{H}}{2M_0 H} = \sin\gamma[\sin\alpha \cos(\theta - \delta) + \sin\beta \cos(\varphi - \delta)] + \cos\gamma(\cos\alpha + \cos\beta) \quad (3)$$

For $H = 0$, the DM interaction forces the magnetic vectors to stay in the **xz** plane with a canting relative to the **x** axis, i.e. $\theta = \varphi = \pi/2$, $\alpha \neq 0$ and $\beta \neq \pi$, resulting in a WFM vector $\vec{M}_{wf} = \vec{M}_1 + \vec{M}_2$ along the **z** axis. By symmetry, $\beta = \pi - \alpha$ and the canting angle $\alpha_0$ can be easily calculated, yielding

$$\tan(2\alpha_0) = \frac{2H_D}{2H_E + H_A} \approx \frac{H_D}{H_E} \quad (4)$$

where the approximation $H_E \gg H_A$ is often experimentally justified.

Magnetization curves $M(H)$ were simulated by minimizing the energy function $u(\alpha, \theta, \beta, \varphi)$ by the gradient method, for $H$ values increasing from a negative $H_{min}$ to a positive $H_{max}$ and back to $H_{min}$. $H$ was changed by small increments (i. e. $H_{i+1} = H_i + \Delta H$) to ensure that a series of contiguous local minima was traced out on the energy surface. The starting fields ($H_{min}$, $H_{max}$) were chosen sufficiently close to saturation to give a single minimum. In this way, the appearance of a second minimum was reflected by the onset of hysteresis in the $M(H)$ curve.

Magnetization loops were initially simulated for the special configurations in which $H$ was applied along the easy axis ($\gamma = 0$), along the DM axis ($\gamma = \pi/2$, $\delta = 0$), and perpendicular to both ($\gamma = \pi/2$, $\delta = \pi/2$). As can be expected for this highly anisotropic system, each of these situations yields a different magnetic response. Then, based on these single-crystal simulations, the main features of the magnetization curve for a polycrystalline sample can be more readily understood.

**(a) *H* parallel to the easy axis (x axis)**

For this configuration, the magnetic vectors remain in the **xz** plane since the torque exerted by $H$ has no component parallel to **x**. The vector product $\vec{M}_1 \times \vec{M}_2$ is antiparallel to vector $\vec{D}$, thus $u_D = H_D \sin(\alpha - \beta) \geq 0$. Some simulated $M(H)$ curves are displayed in Fig. 4.

Fig. 4(a) shows the effect of the exchange field $H_E$ on the magnetization curves, keeping $H_D$ and $H_A$ constant. There is a rather well-defined saturation field $H_{sat}$ which increases for increasing $H_E$, leading to a decrease in the slope of the linear part of $M(H)$. The fact, mentioned earlier, that the measured magnetization is far below saturation even for applied fields as large as 170 kOe, is an indication that $H_{sat}$ is very high for this material, implying $H_E$ values of order $10^6$ Oe.

The behavior is essentially insensitive to changes in $H_D$ and $H_A$, as illustrated by the superposing simulations in Fig. 4(b). However, detailed examination at low fields ($H < 1$ kOe) reveals some subtle differences. As Fig. 4(c) illustrates, the initial part of the magnetization curve is not linear, reflecting a smooth, initially slow rotation of the resultant magnetization vector toward the **x** axis. In addition, for $H_A$ values above a certain threshold, the final approach to the linear $M(H)$ behavior involves a small discontinuous jump, as can be seen for $H_A = 120$ and 200 kOe. This transition, which can be described as the magnetization vector jumping from some intermediate angle to the **x** axis, is closely related to the spin-flop (SF) transition in simple antiferromagnets. For the latter systems there is a limiting angle $\alpha_L$ between the magnetic moments and the applied field, above which the SF does not take place for any field; this angle is given by [14]

$$\alpha_L = \frac{H_A}{2H_E}. \qquad (5)$$

If a DM interaction is also present, it can be surmised that the canting angle $\alpha_0$ given by Eq. 4 should play the role of $\alpha_L$. Thus, combining Eqs. 4 and 5 one obtains $H_A \approx H_D$ as the minimum anisotropy field for the transition to occur. In order to test this idea, we have simulated magnetization curves for various combinations of $H_A$ and $H_D$, with $H_E = 5600$ kOe. Fig. 5 is a ($H_A$, $H_D$) diagram in which open and closed symbols represent combinations for which the SF-like jump does or does not occur, respectively. As can be seen, the boundary between the two kinds of behavior agrees well with the $H_A = H_D$ line.

FIGURE 5

For extremely small values of the $H_A/H_D$ ratio (e.g. $H_A < 1$ kOe when $H_D = 150$ kOe), it is observed that the WFM moment remains locked to the **x** axis when $H\rightarrow 0$, requiring a negative field to be reversed. Remanence and coercive field are very small, as can be seen in Fig. 6.

FIGURE 6

**(b) *H* parallel to the DM vector (y axis)**

Fig. 7(a) exhibits a typical magnetization curve calculated with $H_E = 5600$ kOe, $H_A = 0.1$ kOe and $H_D = 100$ kOe. In this configuration $\vec{H}$ produces a torque on $\vec{M}_1$ and $\vec{M}_2$ resulting in a reconfiguration of all angles, as can be seen in Figs. 7(b) and 7(c). For large and positive $H$, $\theta$ and $\varphi$ are close to 0, in accordance with the saturation along the **y** axis. When the $H$ intensity decreases, the antiferromagnetic exchange interaction forces the angular opening of vectors $\vec{M}_1$ and $\vec{M}_2$. This tendency is enhanced by the DM interaction, at the same pushing the vectors slightly out of the **xy** plane (cf. the inset in Fig. 7(c); at zero field the canting angle is $\alpha \approx 0.5°$ according to Eq. 4). The overall behavior is similar to that of a simple antiferromagnet with a field applied perpendicular to the easy axis. There is no hysteresis for this configuration.

FIGURE 7

### (c) *H* parallel to the z axis

In this configuration the magnetic vectors are in the **xz** plane ($\theta = \phi = \pi/2$) for any applied field. Fig. 8(a) shows full magnetization curves for various $H_D$ values; as for the **x**-axis case, the gross behavior is insensitive to $H_D$. The small-field region, however, reveals hysteresis, with both remanence and coercive field depending on $H_D$ as shown in Fig. 8(b). The remanence is directly associated to the canting which results from the DM interaction.

FIGURE 8

Another characteristic feature of these results is the occurrence of a first-order transition at $H = H_c$. At this point, the sublattice magnetization vectors flip over in such a way that the vector product $\vec{M}_1 \times \vec{M}_2$ changes sign. Before the transition (*e.g.* $H < H_c$ for increasing fields) the DM energy term $u_D$ is positive although the total magnetic energy corresponds to a local minimum, keeping the system locked in that vector configuration. After the transition, $u_D$ is negative and contributes to stabilize a new configuration with positive magnetization. Our simulations indicate that $H_c$ depends strongly on both $H_D$ and $H_A$ and could possibly be a power function of the $H_A/H_D$ ratio, as is suggested by the systematic shown in Fig. 8(c).

## IV. POLYCRYSTALLINE SAMPLE

A numerical simulation such as those discussed in the preceding section yields a function $M(H;\gamma,\delta)$ which represents the magnetization projected on the field, for a given field direction specified by angles ($\gamma,\delta$). Therefore, for a sample made up of randomly oriented crystallites, the resulting magnetization must be calculated by angular integration (limited to the $z \geq 0$ hemisphere to avoid cancellation by symmetry):

$$\overline{M}(H) = \frac{1}{2\pi} \int_0^\pi d\delta \int_0^\pi M(H;\gamma,\delta) \sin\gamma \, d\gamma \qquad (6)$$

The remanence $M_r = \lim_{H \to 0} \overline{M}(H)$ is a quantity of particular interest since it is directly related to the weak ferromagnetic moment $M_{wf}$. As shown in Section III, in zero applied field the $\vec{M}_{wf}$ vector is aligned along the **z** axis, so its absolute value equals $M_r$ when the field is applied along that axis. For any other field direction, specified by the unit vector $\vec{h}$, $M_r$ is the projection of $\vec{M}_{wf}$ onto $\vec{h}$, i.e.

$$M_r(\gamma,\delta) = \vec{M}_{wf} \cdot \vec{h} = M_{wf} \sin\gamma \sin\delta \qquad (7)$$

and the relationship $\overline{M}_r = M_{wf}/2$ is obtained upon angular averaging.

## V. DISCUSSION

It will be next shown how the parameters $H_E$, $H_D$ and $H_A$ for YFeO$_3$ can be determined from experimental data on a polycrystalline sample. The remanence at $T = 4.2$ K (see Fig. 2) is $M_r = 0.48$ emu/g, hence $M_{wf} = 0.96$ emu/g. The canting angle is thus

$$\alpha_0 \approx \sin\alpha_0 = \frac{M_{wf}}{2M_0} = 13 \text{ mrad.} \qquad (8)$$

This result provides the ratio between $H_D$ and $H_E$, according to Eq. 4. In order to determine the exchange field $H_E$, we note that the experimental $M$ vs. $H$ data of Fig. 2 show a linear behavior for fields larger than $H_{irr}$, i.e. the field above which the loop is closed. This is in accordance with our simulations in the high-field regime for the three principal directions (see Figs. 4(a), 7(a) and 8(a), respectively). The simulations have also shown that the saturation field for these directions was very closely given by $H_{sat} \approx 2H_E$ or, equivalently, the differential susceptibility was

$$\chi_{high\ field} \approx \frac{M_S}{H_E} = \frac{2M_0}{H_E}. \tag{9}$$

This result can be understood by analogy to the case of a simple antiferromagnet: in a field applied perpendicularly to the easy axis the magnetization increases linearly, with a susceptibility $\chi_\perp = 2M_0/(2H_E - H_A)$, while for a parallel field the same slope sets in after the spin-flop transition [14]. As will be seen below, $H_E >> H_A$ in the present case.

Eq. 9 has been confirmed by a number of simulations made for moderate fields, at various directions other than the principal ones. It thus appears that a range of applied fields exists for which the magnetic energy is dominated by $H$ and $H_E$, in such a way that anisotropic effects arising from the $H_D$ and $H_A$ terms become negligible. Inspection of the experimental curve in Fig. 2 suggests that $H_{irr}$ ~ 80 kOe is the minimum field for such isotropic behavior. Accordingly, we identify the high-field slope ($\chi$ = 1.29×10$^{-2}$ emu/g·kOe) with Eq. 9 and obtain $H_E \approx 5590$ kOe. Combining Eqs. 4 and 8, the DM parameter $H_D \approx 149$ kOe is obtained. These values are in good agreement with literature ones for YFeO$_3$ (see Table I).

While the remanence is determined only by $H_E$ and $H_D$, the coercive field is a sensitive function of the anisotropy field $H_A$. According to our simulations, a small hysteresis appears for $H$ along the easy-axis direction (Fig. 6), but this is a very small effect, hardly perceptible in comparison to the large coercivities (on the order of tens of kOe) that arise when $H$ is along the **z** axis (Fig. 8(b)). It must be noted, however, that simulated loops such as the latter cannot be directly compared to experimental data for a polycrystalline sample. Indeed, other directions at moderate angles to the **z** axis also contribute, with smaller $H_c$ values and a rounded magnetization reversal rather than a discontinuous jump. Thus, the calculation for $H \| \mathbf{z}$ only gives an upper limit for $H_c$. We have simulated full magnetization loops for a polycrystal, with fixed $H_E$ and $H_D$ and varying $H_A$. Some examples are shown in Fig. 9, from which the sensitivity to $H_A$ can be appreciated. The best agreement with our experimental data was obtained for $H_A \approx 0.5$ kOe. For the set of parameters given in Table I, the whole simulated curve agrees well with the experimental one.

FIGURE 9

Regarding the coexistence of high-field ($H > 1$ kOe) and low-field ($H < 1$ kOe) regimes observed in the minor hysteresis loop of a polycrystalline $YFeO_3$ sample, such as reported in [11] (see Fig. 1), it can be explained in terms of the intrinsic magnetic properties of the $YFeO_3$ system without any impurity contribution.

The high-field regime is the only one present in the magnetization loop measured up to 170 kOe (Fig. 2) for the sample used in this work. According to our simulations, it is directly related to the first-order transition occurring for $H$ applied along the **z** axis, as seen in Fig. 8(b). This transition is promoted by the DM interaction, and the corresponding transition field was shown to be empirically correlated to the ($H_A/H_D$) ratio. This correlation is displayed in Fig. 8(c). From the $H_D$ and $H_A$ values obtained in this work (entered in Fig. 8(c) as an open triangle) a transition field $H_c = 81$ kOe can be estimated. As was discussed in the preceding paragraph, this value is an upper limit to the coercive field of a polycrystalline sample, due to the presence of other local orientations which necessarily have a smaller $H_c$ value. Thus, the measured coercive field of our sample, 22 kOe, can be considered to be in good order-of-magnitude agreement with the calculated value.

The low-field regime, typified by the narrow hysteresis loop seen in Fig. 1, necessarily has a different origin. Considering the order of magnitude of the fields involved, it is most plausible to associate this behavior to the simulated loops shown in Fig.

6, which occur for *H* applied parallel or close to the anisotropy axis (i.e. $\gamma \approx 0$). In the angular averaging defined by Eq. 6, however, such directions make an almost negligible contribution owing to the $\sin(\gamma)$ weighting factor. Therefore, no low-field hysteresis should be apparent for a randomly distributed polycrystalline sample, which seems to be the case for the $YFeO_3$ sample used in this work. The fact that it was observed in the magnetization loop of Fig. 1 might be explained by some texture in the sample of Ref. 11, which may have been induced by magnetocrystalline anisotropy during sample preparation.

In conclusion, the present work strongly suggests that the peculiar properties of $YFeO_3$, such as a high coercivity field, a high irreversibility field and the peculiar shape of the hysteresis loop, are intrinsic properties of this system, fully explainable by its weak-ferromagnetic character.

**Acknowledgments**

This work was supported by Fundação de Amparo à Pesquisa do Estado de São Paulo (FAPESP), Volkswagen Foundation, and Conselho Nacional de Desenvolvimento Científico e Tecnológico (CNPq). The authors are grateful to Prof. Nei F. de Oliveira Jr. for the VSM measurements up to 170 kOe and to Prof. Roberto D. Zysler for a critical reading of the manuscript.

FIGURE CAPTIONS

Figure 1 - Minor hysteresis loop of a YFeO$_3$ polycrystalline sample, measured at $T = 10$ K and up to fields (0.5, 2, 5, 10, 30 and 70 kOe) lower than the irreversibility field (~ 80 kOe). Inset shows the low-field regime of the magnetization curve. (Taken from Ref. 11)

Figure 2 - Experimental (open circles) and simulated (solid line) hysteresis loops of polycrystalline YFeO$_3$. Experimental hysteresis loop was measured up to 170 kOe at 4.2 K. The simulated one was evaluated taking $H_E$ = 5590 kOe, H$_D$ = 149 kOe and H$_A$ = 0.5 kOe.

Figure 3 - Schematic representation of the applied field $\vec{H}$, DM vector $\vec{D}$, and sublattice magnetization vectors $\vec{M}_1$ and $\vec{M}_2$.

Figure 4 - Simulated magnetization loops with $H$ parallel to **x**-axis ($\gamma = 0$): (a) for different $H_E$ values and $H_D$ = 100 kOe, $H_A$ = 0.2 kOe; (b) for different $H_D$ and $H_A$ values and $H_E$ = 5600 kOe; (c) the low-field region for different $H_A$ values and $H_E$ = 5600 kOe, $H_D$ = 100 kOe.

Figure 5 - $H_A$ vs. $H_D$ diagram for the simulated spin-flop-like transition for $\gamma = 0$, with $H_E$ = 5600 kOe. Each point represents a set of ($H_D$, $H_A$) values for which the transition either does (open squares) or does not (solid squares) occur. The dashed line is a freely drawn borderline, close to $H_D = H_A$.

Figure 6 - Simulated hysteresis loop with $H$ parallel to **x** axis ($\gamma = 0$) with $H_E = 5600$ kOe, $H_D = 150$ kOe, and $H_A = 1, 5, 10$ and $1000$ Oe. Inset shows in detail the hysteresis for $H_A = 1$ kOe.

Figure 7 - (a) Simulated hysteresis loop with $H$ parallel to axis **y** ($\gamma = \pi/2$, $\delta = 0$) with $H_E = 5600$ kOe, $H_D = 100$ kOe, and $H_A = 0.2$ kOe; (b) $\alpha$ and $\beta$ vs. $H$; (c) $\theta$ and $\varphi$ vs. $H$ (inset: low-field region).

Figure 8 - (a) Simulated magnetization loops for $H$ parallel to **z** axis ($\gamma = \delta = \pi/2$), with $H_E = 5600$ kOe, $H_A = 0.2$ kOe, and different $H_D$ values (b) Low-field region of the same curves. (c) Transition field $H_C$ vs. $(H_D/H_A)$; open triangle corresponds to the values of $H_A$ and $H_D$ of our polycrystalline sample given in Table I.

Figure 9 - Simulated hysteresis loop for polycrystalline YFeO$_3$ with $H_E = 5590$ kOe, $H_D = 150$ kOe, and different $H_A$ values.

Figure 1

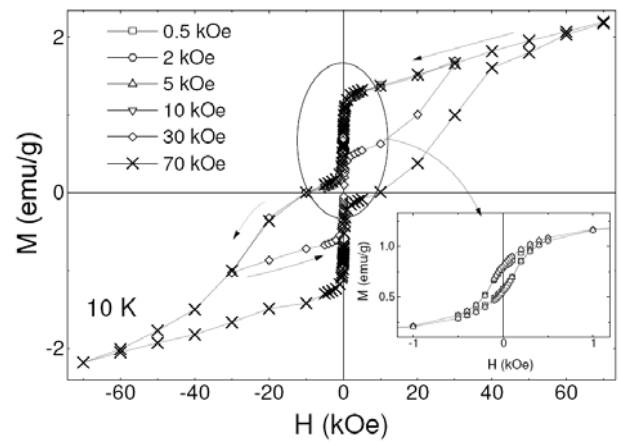

Figure 2

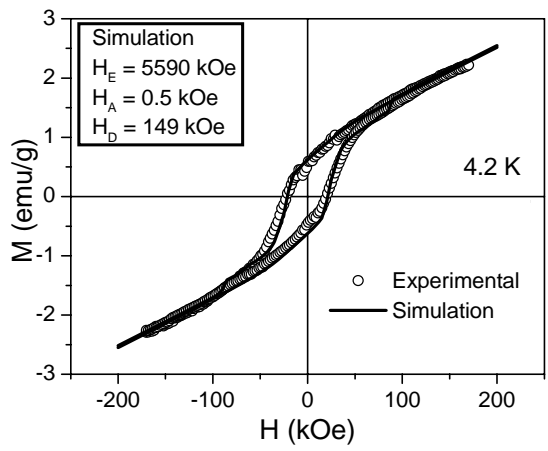

Figure 3

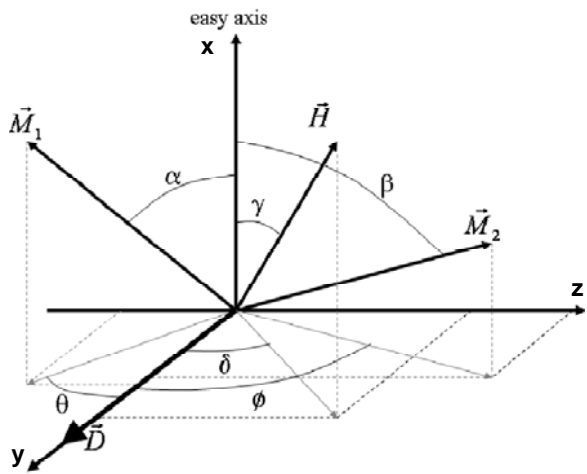

Figure 4

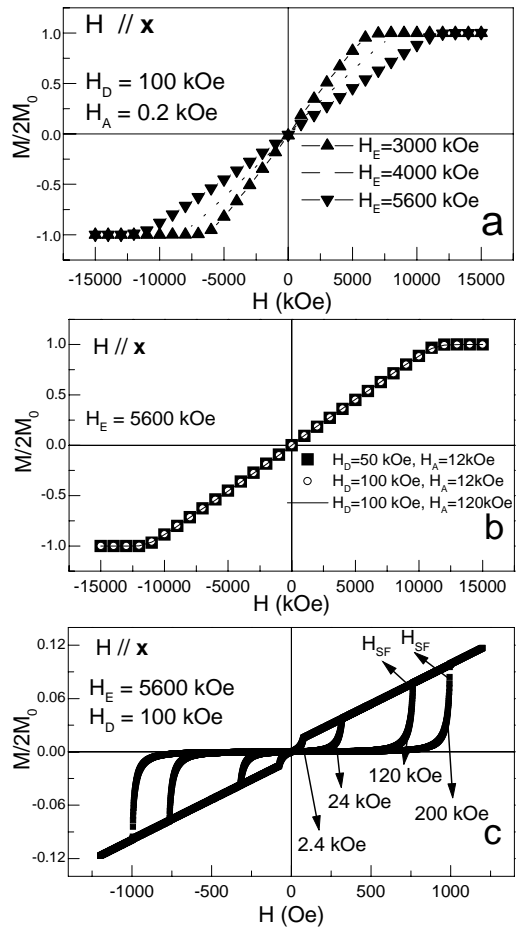

Figura 5

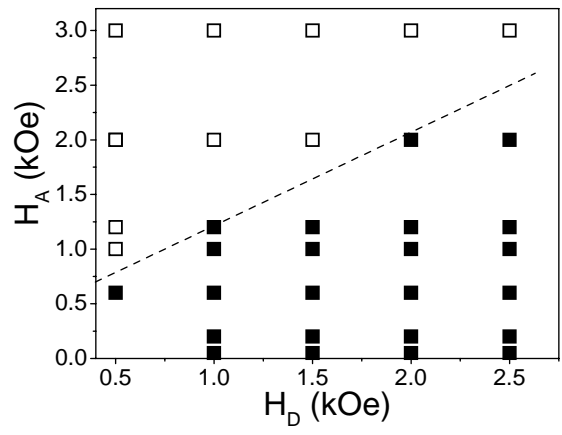

Figura 6

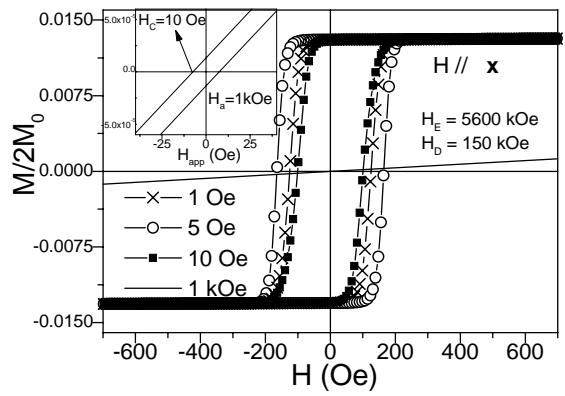

Figure 7

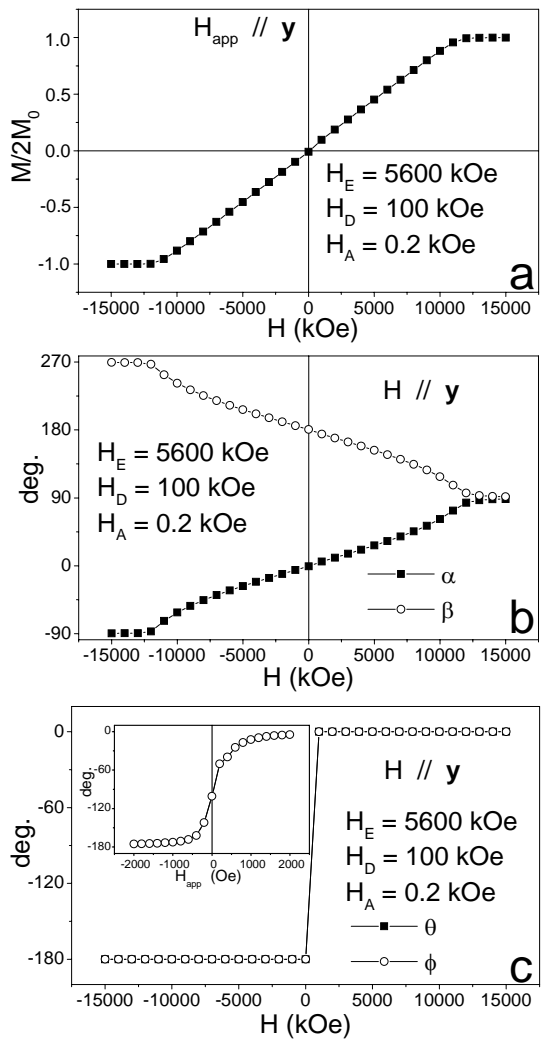

Figure 8

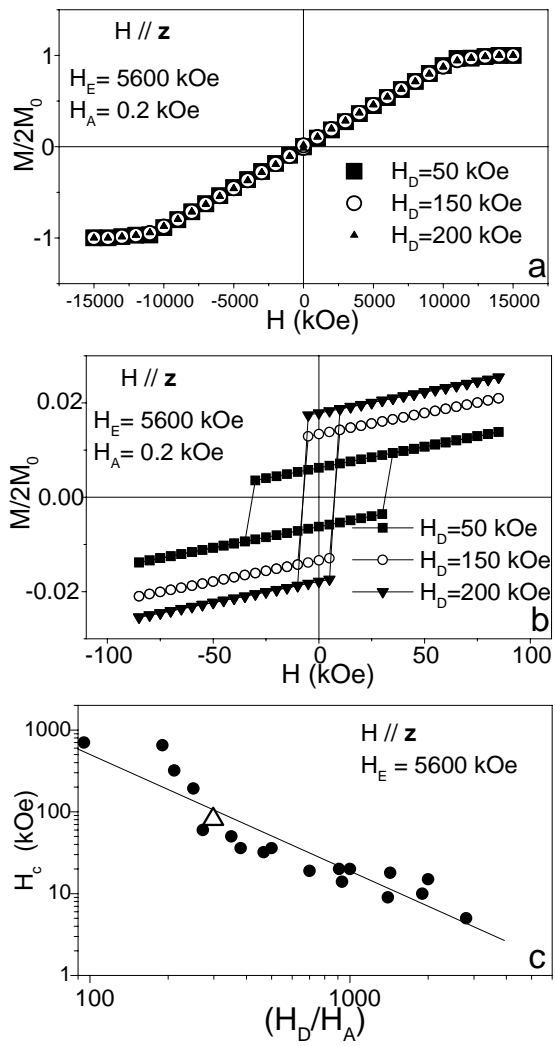

Figura 9

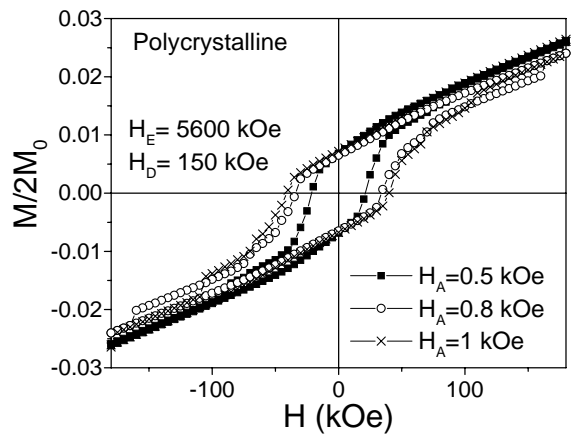

TABLE I – Comparison among the values of the canting angle $\alpha_0$, exchange field $H_E$, Dzyaloshinsky-Moriya field $H_D$ and anisotropy field $H_A$ obtained in the present work and those reported in the literature.

|  | Present work | Ref. [2] | Ref. [5] | Ref. [12] |
|---|---|---|---|---|
| $\alpha_0$ (mrad) | 11 | 8.7 | 8.5 | - |
| $H_E$ (kOe) | 5590 | $6.4 \times 10^3$ | $5.6 \times 10^3$ | $6 \times 10^3$ |
| $H_D$ (kOe) | 149 | $1.4 \times 10^2$ | 95 | $1 \times 10^2$ |
| $H_A$ (kOe) | 0.5 | $2 \times 10^{-1}$ | - | $3.70 \times 10^{-1}$ |